\documentclass[12pt,a4paper,dvips]{article}
\usepackage{a4p}
\usepackage{cite,mcite}
\usepackage{graphicx}
\usepackage{physics}
\usepackage{l3_title}

\journalname{Phys. Lett. B}
\date{July 24, 2000}
\preprint{2000-103}

\def\pb{\mbox{pb$^{-1}$}}

\newcommand{\LL}{\ell^+ \ell^-}
\newcommand{\EE}{\rm e^+ e^-}

\newcommand{\QQ}{\rm q \bar{q}}

\newcommand{\WW}{\rm W^+ W^-}

\newlength{\capindent}
\newlength{\capwidth}
\newlength{\figwidth}
\newcommand{\icaption}[2][!*!,!]{\hspace*{\capindent}%
  \begin{minipage}{\capwidth}
    \ifthenelse{\equal{#1}{!*!,!}}%
      {\caption{#2}}%
      {\caption[#1]{#2}}
  \end{minipage}}

\begin{document}
\begin{titlepage}
\title{Search for a Higgs Boson Decaying into Two Photons in 
$ \bf \mbox{e}^{+}\mbox{e}^{-}$
Interactions at $\bf\sqrt{\bf \mbox{s}}$ = 189 {\bf GeV}}
\begin{Authlist}
{\Large The L3 Collaboration}
\end{Authlist}
%
%
\begin{abstract}
A search is performed for a Higgs boson produced in association with a Z boson 
and decaying into two photons,
using the L3 data collected at LEP at a centre-of-mass energy of 189$\GeV$.
All decay modes of the Z are considered. No signal is observed and 
limits on the branching fraction of the Higgs boson decay 
into two 
photons as a function of the Higgs mass are derived assuming a Standard 
Model production rate. A lower limit on the mass of a  
fermiophobic Higgs is set at 94.9$\GeV$ at 95$\%$ confidence level.
\end{abstract}

\submitted
\end{titlepage}
\section{Introduction}

In the Standard Model, the decay of a Higgs boson h into a photon pair 
occurs 
at the one loop level 
and its branching fraction is small \cite{sm}. 
For example, for Higgs masses in the range 80 $<$ $M_{\mbox{{\scriptsize h}}}$ $<$ 110$\GeV$,
this decay rate lies between 0.1$\%$ and 0.2$\%$.
However, several
extended models predict enhancements of this branching fraction \cite{om}. 
In Two Higgs Doublet Models of Type~I~\cite{thdm}, 
with an appropriate choice of the model parameters,
the lightest CP even Higgs boson does not couple to fermions at tree level.  
Such a Higgs is expected to decay dominantly into a pair of
photons if its mass is below 90$\GeV$ \cite{fphob}.  

\indent
We search for a Higgs boson produced in association with a Z boson through
the process
$\mbox{e}^{+}\mbox{e}^{-}\to$ Zh, followed by the decay h $\to \gamma \gamma$, in
all decay modes of the \mbox{Z} boson.

\indent
These results supersede previous limits obtained by the L3 Collaboration from data 
at lower centre-of-mass energies \cite{l3}.
Similar analyses were published by the OPAL \cite{opal} and the DELPHI \cite{delphi} Collaborations.

\section{Data and Monte Carlo Samples}

The data were collected with the L3 detector \cite{l3det} 
at a centre-of-mass energy 
$\sqrt s$ = 189$\GeV$
and amount to an
integrated luminosity of 176.4 $\pb$.

The Standard Model Higgs production cross section is calculated 
using the HZHA generator \cite{hzha}. 
Monte Carlo samples were
generated using PYTHIA \cite{pythia} for Higgs masses between 50 
and 100$\GeV$. For background studies the following Monte Carlo programs were used:
KK2f \cite{kk2f} ($\EE \to \QQ (\gamma)$), PYTHIA ($\EE \to \rm ZZ$ and $\EE \to \rm Z\EE$),
KORALW \cite{koralw} ($\EE \to \WW$), 
PHOJET \cite{phojet} ($\EE \to \EE \QQ$),
KORALZ \cite{koralz} ($\mbox{e}^{+}\mbox{e}^{-} \to \nu \bar{\nu} (\gamma)$),
GGG \cite{ggg} ($\mbox{e}^{+}\mbox{e}^{-} \to \gamma \gamma (\gamma)$),
BHWIDE \cite{bhwide} ($\mbox{e}^{+}\mbox{e}^{-} \to \mbox{e}^{+}\mbox{e}^{-} (\gamma)$),
TEEGG \cite{teegg} ($\mbox{e}^{+}\mbox{e}^{-} \to \mbox{e}^{+}\mbox{e}^{-} \gamma(\gamma)$), 
DIAG36 \cite{diag36} ($\mbox{e}^{+}\mbox{e}^{-} \to 
\mbox{e}^{+}\mbox{e}^{-}\mbox{e}^{+}\mbox{e}^{-}$) and
EXCALIBUR \cite{excalibur} 
($\mbox{e}^{+}\mbox{e}^{-} \to \mbox{e}^{+}\mbox{e}^{-} \nu \bar{\nu}$). 
The number of simulated events for the most 
important background channels is at least 100 times higher than the number of expected data
events, while this factor is 200 for the expected signal.

The L3 detector response is simulated using the GEANT~3.15
program \cite{geant}, which takes into account the effects of energy loss,
multiple scattering and showers in the detector. GHEISHA \cite{gheisha}
is used to simulate hadronic interactions in the detector.
Time dependent inefficiencies, as monitored during the data taking, are also
simulated.

\section {Analysis Procedures}

A cut-based selection is performed in order to select events with photons
and to identify the Z in its various decay modes.
This gives rise to $\mbox{q} \bar{\mbox{q}} \gamma \gamma $,
$\nu \bar{\nu} \gamma \gamma$ and $\LL \gamma \gamma$, with $\ell =\mbox{e},\mu,\tau$,
final states. The selection criteria for each final state are described
in the following sections and rely on a common photon identification.

Photons are identified as clusters in the electromagnetic calorimeter (BGO)
with an energy greater than 1$\GeV$ and a shower shape compatible
with that of an electromagnetic shower. The ratio of energies
deposited in a 3$\times$3 crystal matrix  
and a 5$\times$5 matrix, centred on the shower axis,
must be greater than 0.95.
The energy deposition in the hadron calorimeter must not exceed 20$\%$ of
the energy deposited in the electromagnetic calorimeter.

In addition, the clusters must not be associated with a charged track within 50 mrad
in the plane perpendicular to the beam axis. 
To suppress photons from initial state radiation,
only photons in the polar angle range 
$45^{\circ}< \theta <135^{\circ}$, $25^{\circ}< \theta <35^{\circ}$ or $145^{\circ}< \theta <155^{\circ}$
are accepted,
which corresponds to the barrel and end-cap regions of the BGO.

In the following selections we require at least two photons.
To ensure that the pair of photons arise from the decay of a heavy resonance
we require the energy of the most energetic photon to be
larger than 10$\GeV$  and the energy of the second most energetic photon to be
larger than 6$\GeV$.

The angular distribution of the di-photon system with respect to the beam direction
is flat for photons coming from the Higgs decay, while it peaks at low polar angles
for those photons originating from double initial state radiation.
Therefore, we require the absolute
value of the cosine of the
polar angle of the di-photon system not to exceed 0.966. 

\subsection{The $\mbox{q} \bar{\mbox{q}}${\boldmath $\gamma \gamma$} final state}

The signature for the $\mbox{q} \bar{\mbox{q}} \gamma \gamma$ final state is a
pair of isolated photons accompanied by two jets. To select these events, we
first apply a hadronic preselection requiring
high multiplicity events. The visible energy normalised to the 
centre-of-mass energy is required to be larger than 0.5 and the energy imbalances
parallel and perpendicular to the beam direction, normalised to the visible energy,
are required to be below 0.4. In order to reduce the background 
from the photon-photon interaction events, we require the energy in a 30$^{\circ}$
cone around the beam pipe to be less than half of the visible energy.
The yield of this preselection is reported in Table \ref{tab_sel1}.

From this sample we select those events which contain at least two photons.
All other particles are clustered in two jets using
the DURHAM jet algorithm \cite{durham}.
To reject photons
coming from neutral hadron decays we require them to be isolated: the energy in a
10$^{\circ}$ cone around the photon direction must be less than 1.5$\GeV$, and in 
a 20$^{\circ}$ cone less than 3.5$\GeV$. The number of charged tracks and calorimeter 
clusters in a 20$^{\circ}$ cone around the photon direction must be below four.
The opening angle between the photons must be larger than 50$^{\circ}$ and
the angle between the photon direction and the nearest jet
must exceed 25$^{\circ}$. 

\begin{figure}[H]
\begin{center}
\begin{tabular}{l}
\includegraphics[width=9cm]{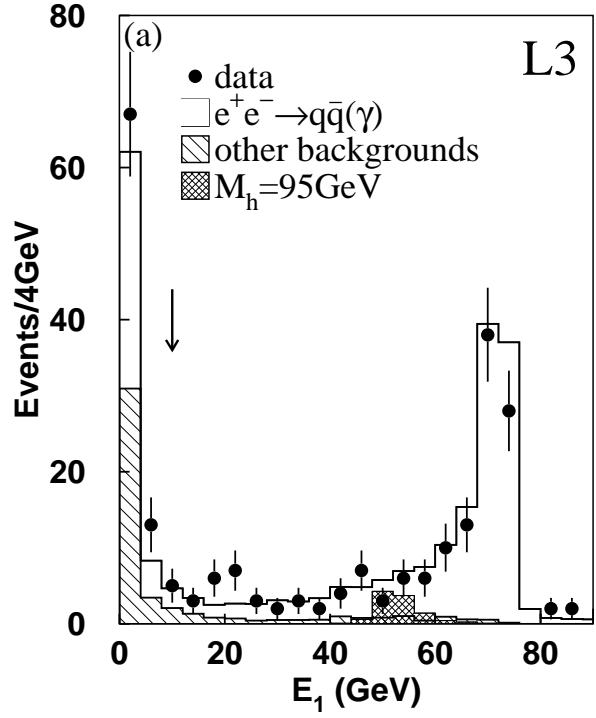} \\
\includegraphics[width=9cm]{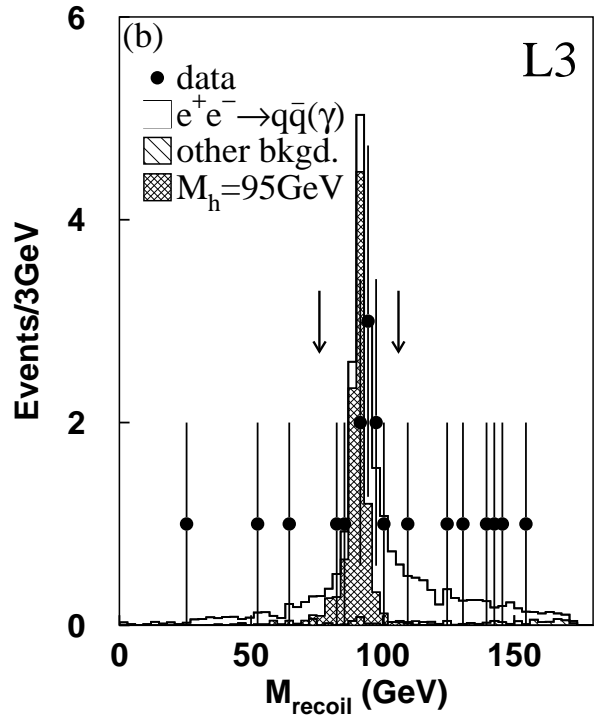} \\
\end{tabular}   
\caption{
Distributions for the $\mbox{q} \bar{\mbox{q}} \gamma \gamma$ final state of 
(a) the energy $\mbox{E}_{1}$ of the most energetic photon
and (b) the recoil mass against the di-photon system    
in data, background and for a 
Higgs boson signal with the mass $M_{\mbox{{\scriptsize h}}}$ =  95$\GeV$.
The signal, corresponding to the Standard Model cross section and
a BR(h$\to \gamma \gamma$) = 1, 
is superimposed and normalised to the integrated luminosity.
}
\label{fig_hadr}
\end{center}
\end{figure}

The energy spectrum of the most energetic photon before any cut is applied on the 
photon energies 
is presented in Figure \ref{fig_hadr}(a).
Figure \ref{fig_hadr}(b) shows the distribution of the recoil mass against the di-photon
system after the selection requirements on the photon energies. 

\indent
We also require the recoil mass against the di-photon system to be consistent with the Z mass,
$|M_{\rm recoil}-M_{\rm Z}|<15$$\GeV$.
This requirement reduces the background from the $\mbox{e}^{+}\mbox{e}^{-} \to \mbox{q}\bar{\mbox{q}}(\gamma)$
process where either a neutral hadron from the Z decay mimics a photon or a photon in the final 
state is emitted; in both cases the recoil
mass against the photons would be smaller than the Z mass. 

The efficiency of this analysis for selecting $\mbox{q} \bar{\mbox{q}} \gamma \gamma$
events is 43$\%$ for a Higgs boson mass of 95$\GeV$. 
The number of selected data and background events are presented in Table \ref{tab_sel1}.  
The dominant
background arises from the process $\mbox{e}^{+}\mbox{e}^{-} \to \mbox{q} \bar{\mbox{q}} (\gamma)$.  

\subsection{The {\boldmath $\nu \bar{\nu} \gamma \gamma$} final state}

The $\nu \bar{\nu} \gamma \gamma$ final state is characterised by the presence of two 
photons and missing energy in the event.
The event selection follows the criteria described in Reference \citen{photons}. 
To reduce the background from the $\mbox{e}^{+}\mbox{e}^{-}\to \gamma \gamma (\gamma)$ process and
from double radiative events with final state particles escaping detection,
we require the photon acoplanarity to be greater than 2.5$^{\circ}$, the total transverse momentum 
of the di-photon system to be greater than 3$\GeV$ and the absolute value of the cosine of the 
polar angle of the missing momentum not to exceed 0.96. 

\begin{figure}[H]
\begin{center}
\begin{tabular}{l}
\includegraphics[width=9cm]{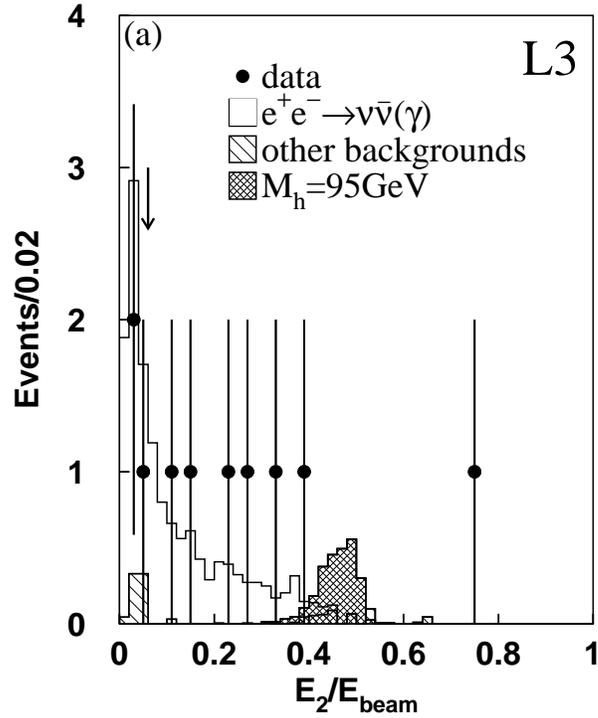} \\
\includegraphics[width=9cm]{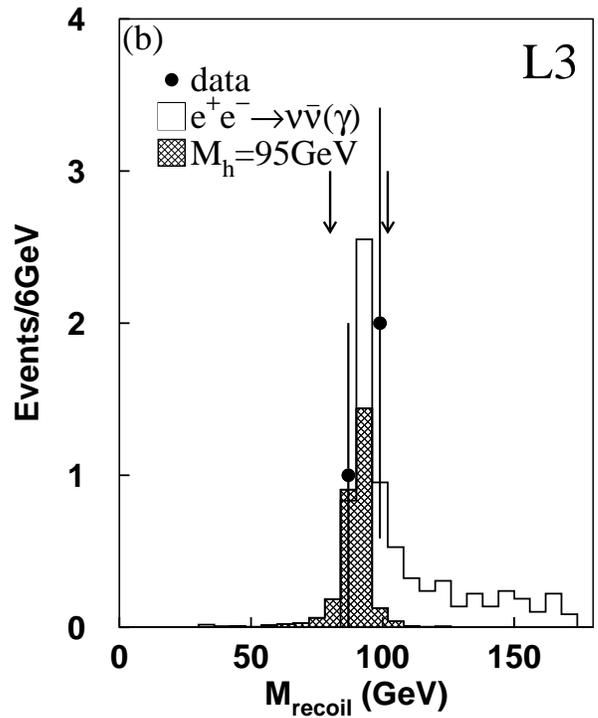} \\
\end{tabular}   
\caption{Distributions for the $\nu \bar{\nu} \gamma \gamma$ final state of 
(a) the energy of the second most energetic photon normalised
to the beam energy
and (b) the recoil mass against the two photons
in data, background and for a
Higgs boson signal with the mass $M_{\mbox{{\scriptsize h}}}$ =  95$\GeV$.
The signal, corresponding to the Standard Model cross section and
a BR(h$\to \gamma \gamma$) = 1, 
is superimposed and normalised to the integrated luminosity.
}
\label{fig_miss}
\end{center}
\end{figure}

The energy of the second most energetic photon 
normalised to the beam energy, $\mbox{E}_{2}/\mbox{E}_{\rm beam}$, is presented 
in Figure \ref{fig_miss}(a). Figure \ref{fig_miss}(b) shows the recoil mass against the two most 
energetic photons after the selection requirements on the photon energies and the di-photon polar 
angle.

Furthermore, the recoil mass must be consistent with the Z boson mass within $\pm$10 GeV. 
The numbers of data and expected background events left after this selection is applied are listed 
in Table \ref{tab_sel2}.  
The signal efficiency 
is 34$\%$ for a Higgs boson of  95$\GeV$ mass. 
The background is due to the
$\mbox{e}^{+}\mbox{e}^{-}\to \nu \bar{\nu} (\gamma)$ process.

\subsection{The {\boldmath $\LL $}{\boldmath $\gamma \gamma$} final state}

The $\LL \gamma \gamma$ final state is characterised by the
presence of two photons and a pair of same type leptons in the
event. First low multiplicity events with a photon pair and a lepton pair are preselected.
 
Electrons are identified as clusters in the BGO with an energy greater than
3$\GeV$ and a matched track. The energy deposited in the hadron
calorimeter must be consistent with the tail of an electromagnetic shower. Moreover 
there must be less than 3$\GeV$ energy deposited in the BGO 
in a 10$^{\circ}$ cone around the electron direction.
 
Muons must have a reconstructed track in the muon chambers with a miss distance to the interaction vertex in
the $r$ - $\phi$ plane smaller than 300 mm and a momentum greater than 3$\GeV$.
The energy in a 10$^{\circ}$ cone around the muon direction must not
exceed 3$\GeV$.
Also events with one muon and one minimum ionising particle in the calorimeters are accepted. 
Background events from cosmic rays are removed by requiring at least one hit in the scintillation counters in a   $\pm$5~ns
time window around the beam crossing time.

Taus are identified as jets 
with one or three tracks in
a 10$^{\circ}$ cone with an energy greater  than 3$\GeV$. The energy in the 
10$^{\circ}$ $-$
30$^{\circ}$ 
cone must not exceed
30\% of the energy in the 0$^{\circ}$ $-$ 10$^{\circ}$ cone around the tau 
direction. 
In order to maintain a high efficiency, events with only one identified tau are also accepted.

The energy of the most energetic lepton
is required to be less than 80$\GeV$ to further reject double radiative di-lepton events.
The result of this preselection is reported in Table \ref{tab_sel3}.

\begin{figure}[H]
\begin{center}
\begin{tabular}{l}
\includegraphics[width=9cm]{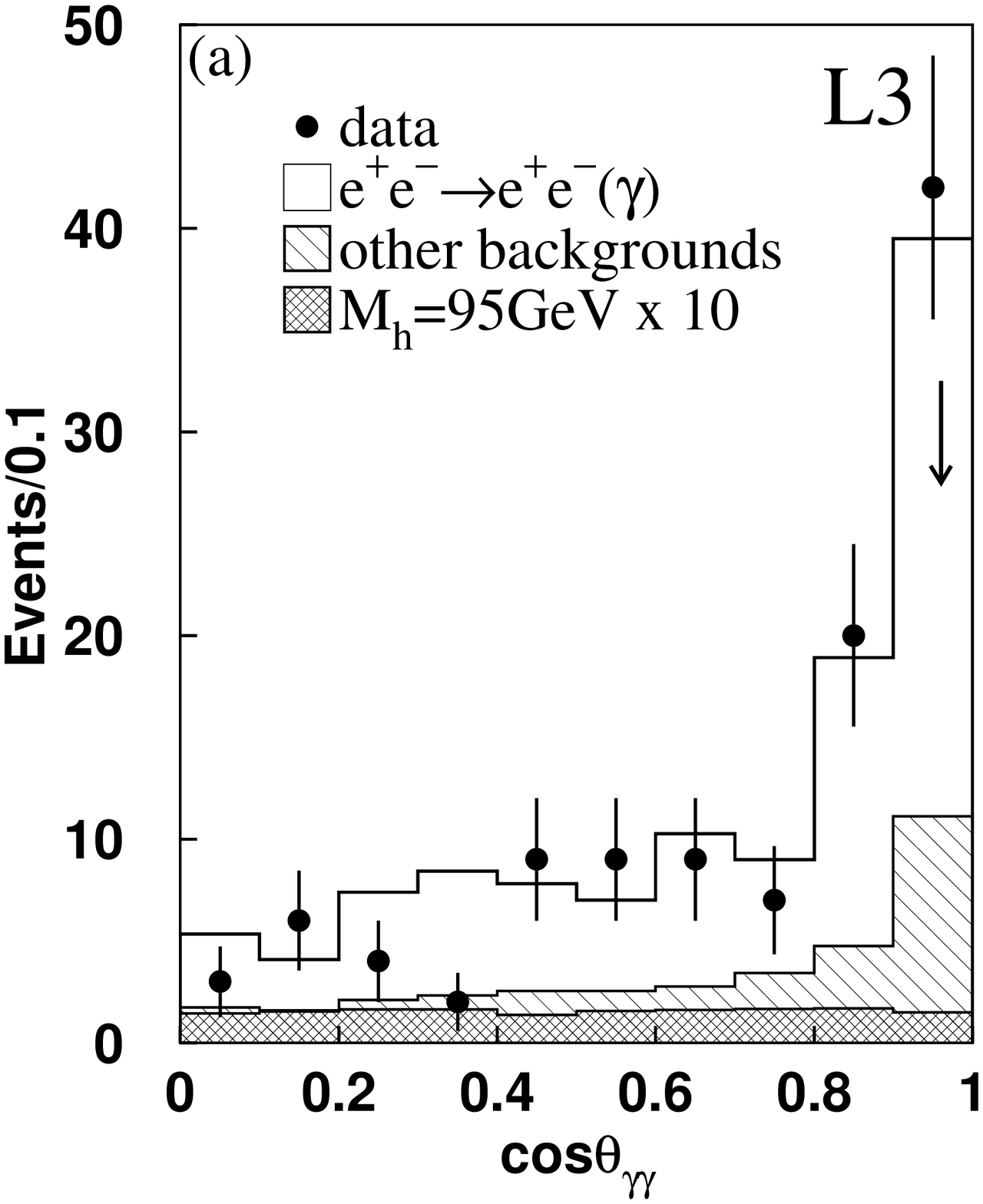} \\
\includegraphics[width=9cm]{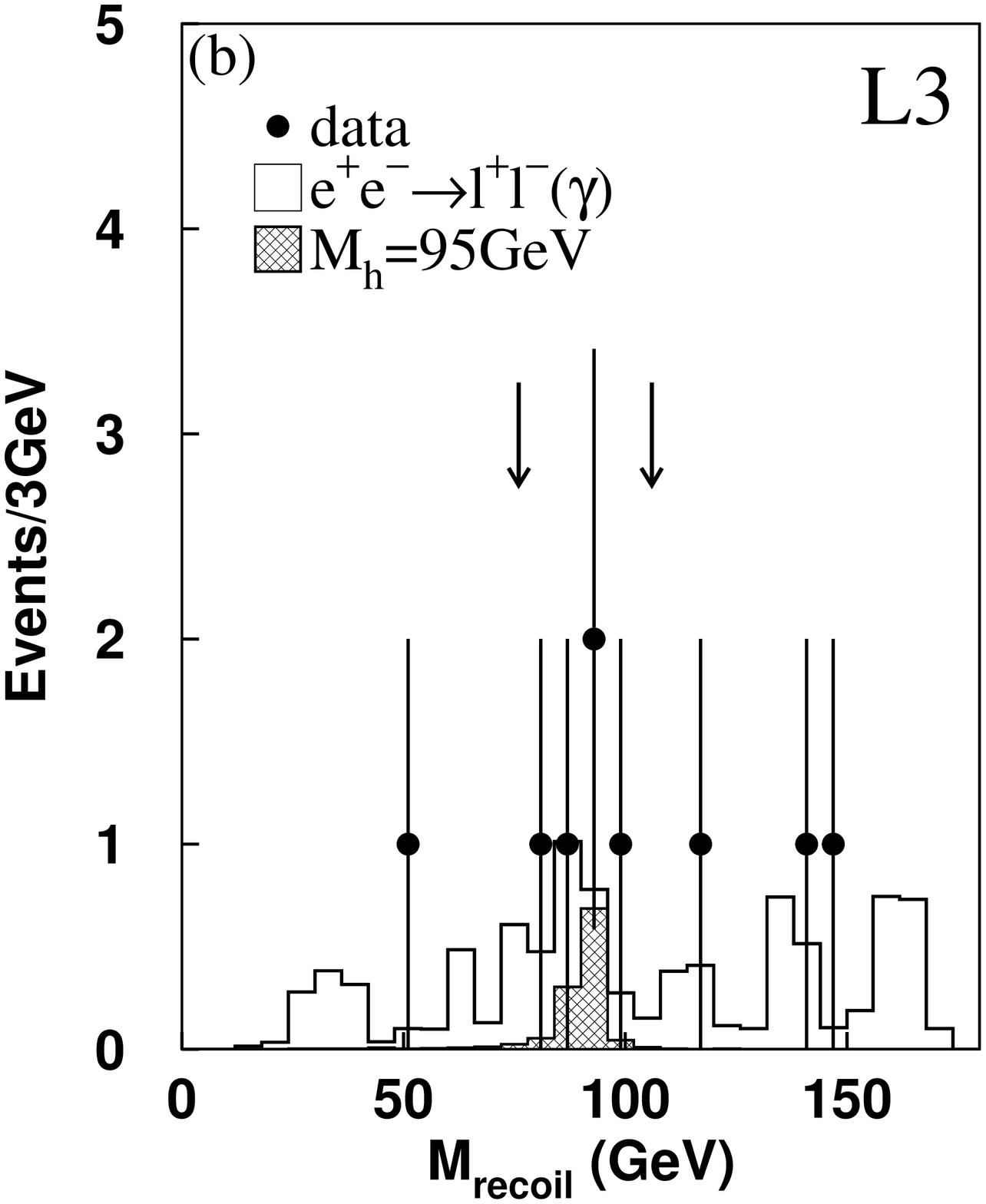} \\
\end{tabular}   
\caption{Distributions for the $\LL \gamma \gamma$ final state of
(a) the cosine of $\theta_{\gamma \gamma}$ for the di-photon system after the preselection and
(b) the recoil mass against the two photons
in data, background and for a
Higgs boson signal with the mass $M_{\mbox{{\scriptsize h}}}$ =  95$\GeV$.
The signal, corresponding to the Standard Model cross section and
a BR(h$\to \gamma \gamma$) = 1,
is superimposed and normalised to the integrated luminosity. A further scale factor of 10
is applied in (a).
}
\label{fig_lept}
\end{center}
\end{figure}

The distribution of the cosine of the polar angle, $\cos \theta_{\gamma \gamma}$,
of the di-photon system is shown in Figure \ref{fig_lept}(a).
Figure \ref{fig_lept}(b) shows the recoil mass against the photons after the cuts on the
photon energies and on the polar angle of the di-photon system.

In addition, we require the recoil mass 
to be consistent with the Z mass,  
$|M_{\rm recoil}-M_{\rm Z}|
<15$$\GeV$.
The number of events selected in data and expected from background processes
is presented in Table \ref{tab_sel3}. The signal efficiency 
in the lepton channel is 29$\%$ for a
Higgs boson with the mass $M_{\mbox{{\scriptsize h}}}$ =  95$\GeV$.
The main backgrounds are due to radiative di-lepton events. 

\section{Results}

The overall efficiency for selecting Zh events varies between 36$\%$ and 42$\%$ for
Higgs boson masses between 50 GeV and 95 GeV, and drops to 30$\%$ at the
kinematic limit.
 
\begin{figure}[H]
\begin{center}
\begin{tabular}{l}
\includegraphics[width=15cm]{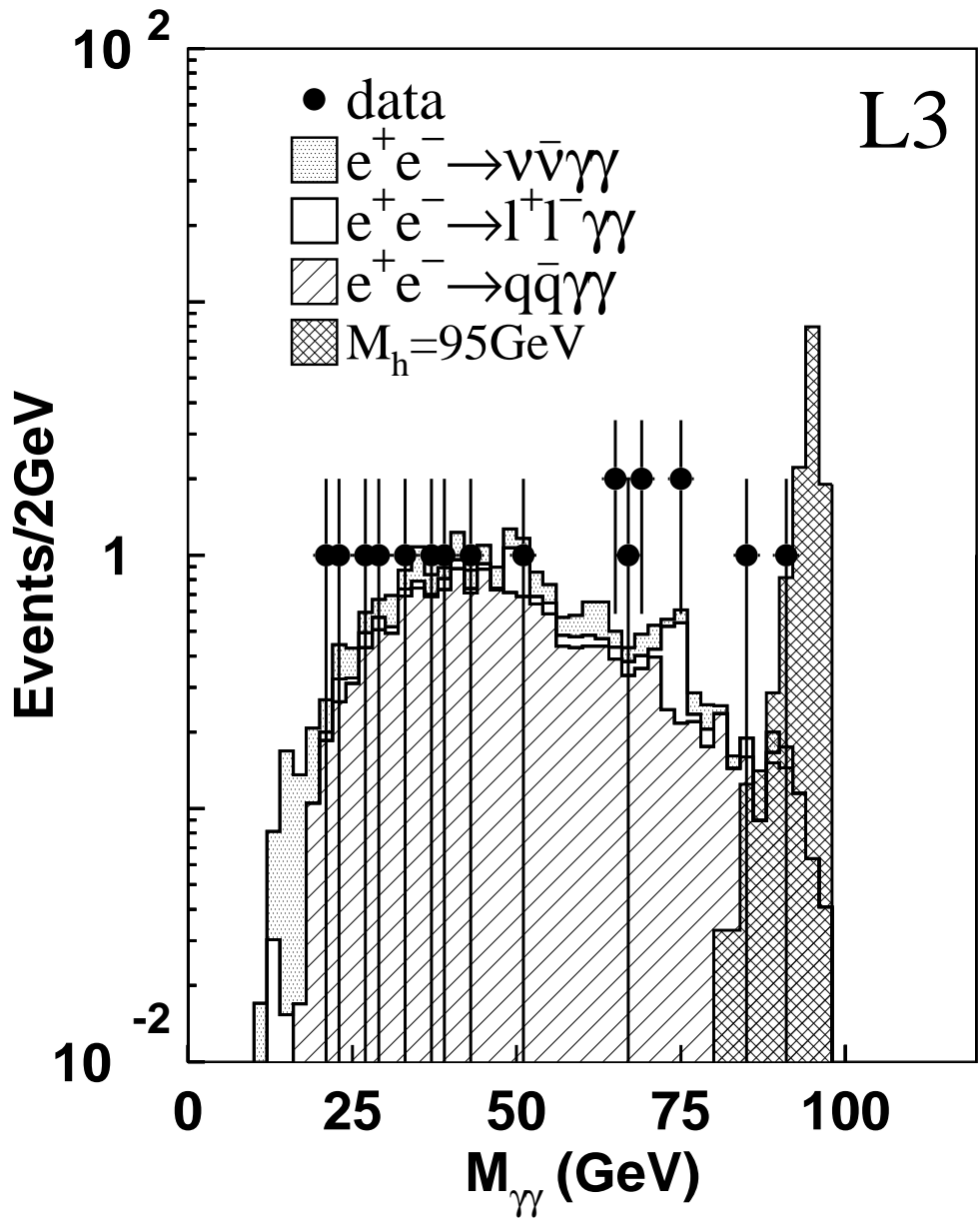} \\
\end{tabular}
\caption{The distribution of the reconstructed di-photon invariant mass  
for all Z final states combined, after the
final selection, in data, background and for a
Higgs boson signal with the mass $M_{\mbox{{\scriptsize h}}}$ =  95$\GeV$.
The signal, assuming the Standard Model cross section and a 
BR(h$\to \gamma \gamma$) = 1, 
is superimposed and normalised to the integrated luminosity.
}
\label{fig_res}
\end{center}
\end{figure}

Since no signal is observed in the data, 
we evaluate the confidence level \cite{cl} for the absence of a Higgs signal 
using the reconstructed di-photon invariant mass as final discriminant variable.
This distribution is shown in Figure \ref{fig_res}
for all Z final states combined.

The calculation of the limits takes into account systematic uncertainties of 1$\%$
from the signal Monte Carlo statistics, 1.5$\%$ from the simulation of the photon 
isolation criteria and 4$\%$ on the number of expected background events. 
The effects from the energy and angular resolution of the 
photons and the systematic uncertainty on the integrated luminosity are found to 
be negligible. 

\begin{figure}[H]
\begin{center}
\begin{tabular}{l}
\includegraphics[width=15cm]{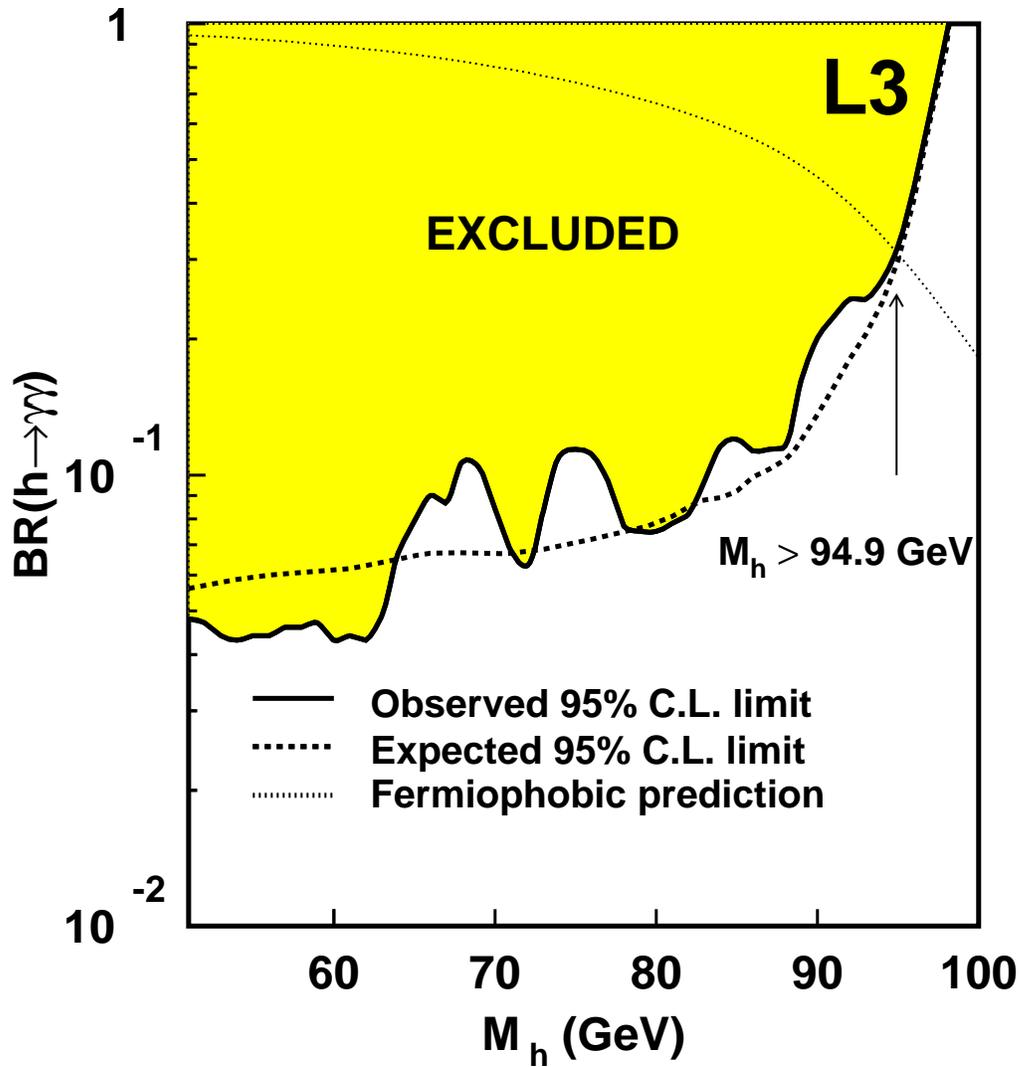} \\
\end{tabular}
\caption{Excluded values of the $\mbox{BR}(\mbox{h}\to \gamma \gamma)$
as a function of the Higgs mass, in the assumption of a Standard Model production cross section.
The expected 95$\%$ confidence level limit and the theoretical prediction are also presented.
}
\label{fig_cl}
\end{center}
\end{figure}

Figure \ref{fig_cl} shows the measured upper limits on the
$\mbox{BR}(\mbox{h} 
\to \gamma \gamma)$ as a function
of the Higgs mass assuming a Standard Model rate for the Zh production,
along with the
expected limits as calculated from a large sample of Monte Carlo experiments. 
The
theoretical prediction is also shown for a fermiophobic Higgs boson as
calculated with the HDECAY program \cite{hdecay}. The observed limit for
$\mbox{BR}(\mbox{h}
\to \gamma \gamma)$ = 1 is 98$\GeV$. 
The lower limit on the mass of a fermiophobic Higgs boson is set at

\begin{center}
$M_{\mbox{{\scriptsize h}}} > 94.9$$\GeV$ at 95$\%$ confidence level.
\end{center}

The expected mass limit is 95.1$\GeV$. 

\section*{Acknowledgements}

We wish to express our gratitude to the CERN accelerator division for the
excellent performance of the LEP machine. We acknowledge with appreciation 
the effort of engineers, technicians and support staff who have participated
in the construction and maintenance of this experiment.

\newpage

\typeout{   }     
\typeout{Using author list for paper 217 -- ? }
\typeout{$Modified: Tue Jul 18 08:24:07 2000 by clare $}
\typeout{!!!!  This should only be used with document option a4p!!!!}
\typeout{   }
%
%
%
%
%
%

\newcount\tutecount  \tutecount=0
\def\tutenum#1{\global\advance\tutecount by 1 \xdef#1{\the\tutecount}}
\def\tute#1{$^{#1}$}
\tutenum\aachen            
\tutenum\nikhef            
\tutenum\mich              
\tutenum\lapp              
\tutenum\basel             
\tutenum\lsu               
\tutenum\beijing           
\tutenum\berlin            
\tutenum\bologna           
\tutenum\tata              
\tutenum\ne                
\tutenum\bucharest         
\tutenum\budapest          
\tutenum\mit               
\tutenum\debrecen          
\tutenum\florence          
\tutenum\cern              
\tutenum\wl                
\tutenum\geneva            
\tutenum\hefei             
\tutenum\seft              
\tutenum\lausanne          
\tutenum\lecce             
\tutenum\lyon              
\tutenum\madrid            
\tutenum\milan             
\tutenum\moscow            
\tutenum\naples            
\tutenum\cyprus            
\tutenum\nymegen           
\tutenum\caltech           
\tutenum\perugia           
\tutenum\cmu               
\tutenum\prince            
\tutenum\rome              
\tutenum\peters            
\tutenum\potenza           
\tutenum\salerno           
\tutenum\ucsd              
\tutenum\santiago          
\tutenum\sofia             
\tutenum\korea             
\tutenum\alabama           
\tutenum\utrecht           
\tutenum\purdue            
\tutenum\psinst            
\tutenum\zeuthen           
\tutenum\eth               
\tutenum\hamburg           
\tutenum\taiwan            
\tutenum\tsinghua          

{
\parskip=0pt
\noindent
{\bf The L3 Collaboration:}
\ifx\selectfont\undefined
 \baselineskip=10.8pt
 \baselineskip\baselinestretch\baselineskip
 \normalbaselineskip\baselineskip
 \ixpt
\else
 \fontsize{9}{10.8pt}\selectfont
\fi
\medskip
\tolerance=10000
\hbadness=5000
\raggedright
\hsize=162truemm\hoffset=0mm
\def\r{\rlap,}
\noindent

M.Acciarri\r\tute\milan\
P.Achard\r\tute\geneva\ 
O.Adriani\r\tute{\florence}\ 
M.Aguilar-Benitez\r\tute\madrid\ 
J.Alcaraz\r\tute\madrid\ 
G.Alemanni\r\tute\lausanne\
J.Allaby\r\tute\cern\
A.Aloisio\r\tute\naples\ 
M.G.Alviggi\r\tute\naples\
G.Ambrosi\r\tute\geneva\
% ...
\newpage
\begin{list}{A}{\itemsep=0pt plus 0pt minus 0pt\parsep=0pt plus 0pt minus 0pt
                \topsep=0pt plus 0pt minus 0pt}
\item[\aachen]
 I. Physikalisches Institut, RWTH, D-52056 Aachen, FRG$^{\S}$\\
 III. Physikalisches Institut, RWTH, D-52056 Aachen, FRG$^{\S}$
\item[\nikhef] National Institute for High Energy Physics, NIKHEF, 
     and University of Amsterdam, NL-1009 DB Amsterdam, The Netherlands
\item[\mich] University of Michigan, Ann Arbor, MI 48109, USA
\item[\lapp] Laboratoire d'Annecy-le-Vieux de Physique des Particules, 
     LAPP,IN2P3-CNRS, BP 110, F-74941 Annecy-le-Vieux CEDEX, France
\item[\basel] Institute of Physics, University of Basel, CH-4056 Basel,
     Switzerland
\item[\lsu] Louisiana State University, Baton Rouge, LA 70803, USA
\item[\beijing] Institute of High Energy Physics, IHEP, 
  100039 Beijing, China$^{\triangle}$ 
\item[\berlin] Humboldt University, D-10099 Berlin, FRG$^{\S}$
\item[\bologna] University of Bologna and INFN-Sezione di Bologna, 
     I-40126 Bologna, Italy
\item[\tata] Tata Institute of Fundamental Research, Bombay 400 005, India
\item[\ne] Northeastern University, Boston, MA 02115, USA
\item[\bucharest] Institute of Atomic Physics and University of Bucharest,
     R-76900 Bucharest, Romania
\item[\budapest] Central Research Institute for Physics of the 
     Hungarian Academy of Sciences, H-1525 Budapest 114, Hungary$^{\ddag}$
\item[\mit] Massachusetts Institute of Technology, Cambridge, MA 02139, USA
\item[\debrecen] KLTE-ATOMKI, H-4010 Debrecen, Hungary$^\P$
\item[\florence] INFN Sezione di Firenze and University of Florence, 
     I-50125 Florence, Italy
\item[\cern] European Laboratory for Particle Physics, CERN, 
     CH-1211 Geneva 23, Switzerland
\item[\wl] World Laboratory, FBLJA  Project, CH-1211 Geneva 23, Switzerland
\item[\geneva] University of Geneva, CH-1211 Geneva 4, Switzerland
\item[\hefei] Chinese University of Science and Technology, USTC,
      Hefei, Anhui 230 029, China$^{\triangle}$
\item[\seft] SEFT, Research Institute for High Energy Physics, P.O. Box 9,
      SF-00014 Helsinki, Finland
\item[\lausanne] University of Lausanne, CH-1015 Lausanne, Switzerland
\item[\lecce] INFN-Sezione di Lecce and Universit\'a Degli Studi di Lecce,
     I-73100 Lecce, Italy
\item[\lyon] Institut de Physique Nucl\'eaire de Lyon, 
     IN2P3-CNRS,Universit\'e Claude Bernard, 
     F-69622 Villeurbanne, France
\item[\madrid] Centro de Investigaciones Energ{\'e}ticas, 
     Medioambientales y Tecnolog{\'\i}cas, CIEMAT, E-28040 Madrid,
     Spain${\flat}$ 
\item[\milan] INFN-Sezione di Milano, I-20133 Milan, Italy
\item[\moscow] Institute of Theoretical and Experimental Physics, ITEP, 
     Moscow, Russia
\item[\naples] INFN-Sezione di Napoli and University of Naples, 
     I-80125 Naples, Italy
\item[\cyprus] Department of Natural Sciences, University of Cyprus,
     Nicosia, Cyprus
\item[\nymegen] University of Nijmegen and NIKHEF, 
     NL-6525 ED Nijmegen, The Netherlands
\item[\caltech] California Institute of Technology, Pasadena, CA 91125, USA
\item[\perugia] INFN-Sezione di Perugia and Universit\'a Degli 
     Studi di Perugia, I-06100 Perugia, Italy   
\item[\cmu] Carnegie Mellon University, Pittsburgh, PA 15213, USA
\item[\prince] Princeton University, Princeton, NJ 08544, USA
\item[\rome] INFN-Sezione di Roma and University of Rome, ``La Sapienza",
     I-00185 Rome, Italy
\item[\peters] Nuclear Physics Institute, St. Petersburg, Russia
\item[\potenza] INFN-Sezione di Napoli and University of Potenza, 
     I-85100 Potenza, Italy
\item[\salerno] University and INFN, Salerno, I-84100 Salerno, Italy
\item[\ucsd] University of California, San Diego, CA 92093, USA
\item[\santiago] Dept. de Fisica de Particulas Elementales, Univ. de Santiago,
     E-15706 Santiago de Compostela, Spain
\item[\sofia] Bulgarian Academy of Sciences, Central Lab.~of 
     Mechatronics and Instrumentation, BU-1113 Sofia, Bulgaria
\item[\korea]  Laboratory of High Energy Physics, 
     Kyungpook National University, 702-701 Taegu, Republic of Korea
\item[\alabama] University of Alabama, Tuscaloosa, AL 35486, USA
\item[\utrecht] Utrecht University and NIKHEF, NL-3584 CB Utrecht, 
     The Netherlands
\item[\purdue] Purdue University, West Lafayette, IN 47907, USA
\item[\psinst] Paul Scherrer Institut, PSI, CH-5232 Villigen, Switzerland
\item[\zeuthen] DESY, D-15738 Zeuthen, 
     FRG
\item[\eth] Eidgen\"ossische Technische Hochschule, ETH Z\"urich,
     CH-8093 Z\"urich, Switzerland
\item[\hamburg] University of Hamburg, D-22761 Hamburg, FRG
\item[\taiwan] National Central University, Chung-Li, Taiwan, China
\item[\tsinghua] Department of Physics, National Tsing Hua University,
      Taiwan, China
\item[\S]  Supported by the German Bundesministerium 
        f\"ur Bildung, Wissenschaft, Forschung und Technologie
\item[\ddag] Supported by the Hungarian OTKA fund under contract
numbers T019181, F023259 and T024011.
\item[\P] Also supported by the Hungarian OTKA fund under contract
  numbers T22238 and T026178.
\item[$\flat$] Supported also by the Comisi\'on Interministerial de Ciencia y 
        Tecnolog{\'\i}a.
\item[$\sharp$] Also supported by CONICET and Universidad Nacional de La Plata,
        CC 67, 1900 La Plata, Argentina.
\item[$\diamondsuit$] Also supported by Panjab University, Chandigarh-160014, 
        India.
\item[$\triangle$] Supported by the National Natural Science
  Foundation of China.
\end{list}
}
\vfill


\newpage

\newpage
\begin{table} [ht]
\begin{center}
\hspace*{-1.cm}
\begin{tabular}{|c||c||c||c|c|c|c|}\hline\hline
                  &Data &$\Sigma$Bkg. &$\mbox{q}\bar{\mbox{q}}(\gamma)$& WW& 
$\mbox{Z}\mbox{e}^{+}\mbox{e}^{-}$& 
ZZ\\
\hline
\hline
Preselection
                              &8146&8221.7&5745.9&2309.8&58.2&109.8 \\
\hline
Selection
 &\phantom{0}\phantom{0}10&\phantom{0}\phantom{0}16.2&\phantom{0}
\phantom{0}16.0&\phantom{0}\phantom{0}\phantom{0}0.0&\phantom{0}0.1&
\phantom{0}\phantom{0}0.1    \\ 
\hline\hline
\end{tabular}
\caption{
Number of events expected from Standard Model processes compared to the observed number
of events, after the preselection and selection steps, for the $\mbox{q}\bar{\mbox{q}} \gamma \gamma$ 
final state.
}
\label{tab_sel1}
\end{center}
\end{table}

\begin{table} [ht]
\begin{center}
\hspace*{-1.cm}
\begin{tabular}{|c||c||c||c|}\hline\hline
                  &Data& $\nu \bar{\nu}(\gamma)$ \\
\hline
\hline
Selection
                              &3&4.3 \\ 
\hline\hline
\end{tabular}
\caption{
Number of events expected from the Standard Model process $\rm e^{+}\rm e^{-} \to \nu \bar{\nu} (\gamma)$ 
compared to the observed number
of events, for the $\nu \bar{\nu} \gamma \gamma$ final state.
}
\label{tab_sel2}
\end{center}
\end{table}

\begin{table} [ht]
\begin{center}
\hspace*{-1.cm}
\begin{tabular}{|c||c||c||c|c|c|c|}\hline\hline
                  &Data& $\Sigma$Bkg.& $\mbox{e}^{+} \mbox{e}^{-} (\gamma)$& $\mu^{+} \mu^{-}(\gamma)$& 
$\tau^{+} \tau^{-}(\gamma)$& 
4 fermion \\
\hline
\hline
Preselection
                              &86&93.8&66.4&14.1&9.9&3.4 \\
\hline
Selection         
  &\phantom{0}5&\phantom{0}2.5&\phantom{0}1.1&\phantom{0}0.7&0.7&0.0   \\ 
\hline\hline
\end{tabular}
\caption{
Number of events expected from Standard Model processes compared to the observed number
of events, after the preselection and selection steps, for the $\LL \gamma \gamma$ final state.
}
\label{tab_sel3}
\end{center}
\end{table}


\newpage

\newpage

\newpage

\newpage
\newpage
%
\vfill\newpage

\end{document}